# Visualizing the out-of-plane electronic dispersions in an intercalated transition metal dichalcogenide

Xian P. Yang[1,*,†], Harrison LaBollita[2,*], Zi-Jia Cheng[1,*], Hari Bhandari[3,4,*], Tyler A. Cochran[1,*], Jia-Xin Yin[1], Md. Shafayat Hossain[1], Ilya Belopolski[1], Qi Zhang[1], Yuxiao Jiang[1], Nana Shumiya[1], Daniel Multer[1], Maksim Liskevich[1], Dmitry A. Usanov[5], Yanliu Dang[6,7], Vladimir N. Strocov[5], Albert V. Davydov[6], Nirmal J. Ghimire[3,4], Antia S. Botana[2], M. Zahid Hasan[1,8,9,†]

[1]Laboratory for Topological Quantum Matter and Advanced Spectroscopy (B7), Department of Physics, Princeton University, Princeton, NJ, USA
[2]Department of Physics, Arizona State University, Tempe, AZ, USA
[3]Department of Physics and Astronomy, George Mason University, Fairfax, VA, USA
[4]Quantum Science and Engineering Center, George Mason University, Fairfax, VA, USA
[5]Swiss Light Source, Paul Scherrer Institute, Villigen, Switzerland
[6]Materials Science and Engineering Division, National Institute of Standards and Technology (NIST), Gaithersburg, MD, USA
[7]Department of Electrical and Computer Engineering, Purdue University, West Lafayette, IN, USA
[8]Princeton Institute for Science and Technology of Materials, Princeton University, Princeton, NJ, USA
[9]Materials Sciences Division, Lawrence Berkeley National Laboratory, Berkeley, CA, USA

Layered transition metal dichalcogenides have rich phase diagram and they feature two dimensionality on numerous physical properties. $Co_{1/3}NbS_2$ is one of the newest members of this family where Co atoms are intercalated into the Van der Waals gaps between $NbS_2$ layers. We study the three-dimensional electronic band structure of $Co_{1/3}NbS_2$ using both surface and bulk sensitive angle-resolved photoemission spectroscopy. We show that the electronic bands don't fit into the rigid-band-shift picture after the Co intercalation. Instead, $Co_{1/3}NbS_2$ displays a different orbital character near the Fermi level compared to the pristine $NbS_2$ compound and has a clear band dispersion in $k_z$ direction despite its layered structure. Our photoemission study demonstrates the out-of-plane electronic correlations introduced by the Co intercalation, thus offering a new perspective on this compound. Finally, we propose how Fermi level tuning could lead to exotic phases such as spin density wave instability.

Transition metal dichalcogenides (TMDs) are a family of interesting materials hosting exotic properties including superconductivity, charge density waves (CDW), Mott insulating phases, and nontrivial topology [1][2][3][4]. Moreover, these layered materials feature various stacking patterns [5] such as 2H and 3R. Many of them are also mechanically exfoliable, enabling the realization of twistronics [6]. A well-known member in this family is 2H-$NbS_2$, a superconductor with a critical temperature around 6 K [7].

Given the fact that these TMDs are layered materials, various types of atoms can be intercalated into the van der Waals gaps. Therefore, intercalated TMDs form a new, large class of materials with novel physical properties. For example, superconducting 2H-$NbS_2$ can be intercalated by Fe and Co atoms to show an antiferromagnetic (AFM) order accompanied by a suppression of superconductivity [8]. Recently, a transport study [9] reports a large anomalous Hall effect (AHE) below Néel temperature ($T_N$), which cannot be explained by the presence of a small ferromagnetic moment. AHE has been well known in ferromagnetic materials with a net magnetic moment and strong spin-orbit coupling (SOC) [10]. Complex antiferromagnetic textures such as non-collinear and non-coplanar orderings are also known to give rise to AHE [11][12][13][14]. Recently, AHE has been proposed in some collinear antiferromagnets as well, mediated by their special arrangement of the magnetic atoms within the crystals [15][16][17]. Otherwise, a collinear antiferromagnet is not expected to display AHE.

The collinear antiferromagnetic structure of $Co_{1/3}NbS_2$ [18] does not support either of the above two criteria for AHE. However, the electronic band structure calculated in the paramagnetic state shows the presence of topological bands close to the Fermi energy and thus it was proposed that either a more complex magnetic structure inaccessible to the diffraction experiment in 1980s or an interplay between the magnetic ordering and the electronic band structure accounts for the large AHE [9]. While the apparent inconsistency with respect to the magnetic structure calls for further studies on the magnetic ground state in this material, photoemission can visualize the band structure, thus serving as a direct way to understand the exotic AHE in $Co_{1/3}NbS_2$ via studying its electronic structure. Moreover, a transport study on exfoliated $Co_{1/3}NbS_2$ crystals shows 60 percent AHE quantization per layer [19]. A natural next step is to seek for a quantized AHE in an AFM. Nontrivial bands could help to materialize such a goal.

Motivated by these exciting results, we conduct angle-resolved photoemission spectroscopy (ARPES) measurements on single crystals of $Co_{1/3}NbS_2$ to understand its band structure. We compare $Co_{1/3}NbS_2$ bands with that of the pristine compound ($2H-NbS_2$) to emphasize that the Co intercalation doesn't fit into the rigid-band-shift picture [8][20][21]. We argue that a crossing from two linear bands is not shown in $NbS_2$, and this feature is a consequence of the Co intercalation. Moreover, polarization dependent ARPES data reveal a different nature of the d orbitals near the Fermi level. Via bulk sensitive soft X-ray ARPES, we then demonstrate that the bands after Co intercalation show a clear dispersion along $k_z$ direction despite the large $c$ lattice constant value. Therefore, Co atoms increases the electronic hybridization between the magnetic Co atoms and metallic $NbS_2$ layers, thus giving rise to a strong band dispersion in the out-of-plane direction. Finally, we compare surface and bulk sensitive ARPES to demonstrate how the surface environment could induce a band splitting not observed in the bulk. Our ARPES study, thus, provides a new perspective on intercalated TMD compounds that could lead to exotic phases such as spin density wave (SDW).

$Co_{1/3}NbS_2$ samples were grown by chemical vapor transport as described elsewhere [9]. High-quality, hexagonal flat single crystals with centimeter size were used for our measurements. Ultraviolet ARPES experiments were carried out at the Bloch beamline in the MAX IV and the I05 beamline of Diamond Light Source. The energy and angle resolution were better than 20 meV and 0.2 degree, respectively. Soft X-ray ARPES was performed at the ADRESS beamline at Switzerland Light Source (SLS) with 66 (71) meV energy resolution at the photon energy of 370 (413) eV. The sample temperature was kept under 20 K below the transition temperature. Samples were cleaved *in situ* under a pressure lower than $5\times10^{-11}$ Torr, producing shiny surfaces.

Electronic structure calculations of bulk $Co_{1/3}NbS_2$ were carried out with density-functional theory (DFT) framework using the all-electron, full potential code WIEN2k [22] based on the augmented plane wave plus local orbital (APW+lo) basis set. The Perdew-Burke-Ernzerhof (PBE) version of the generalized gradient approximation (GGA) [23] was chosen as the exchange correlation functional. A very dense $k$-mesh of $22 \times 22 \times 9$ was used for integration in the Brillouin zone (BZ). A $R_{MT}K_{max}$ of 7.0 was used for all calculations. For the crystal structure, we used the experimentally derived crystal data obtained from single crystal X-ray diffraction (XRD) experiments [9]. Muffin tin radii were 2.47 a. u. for Nb, 2.36 a. u. for Co, and 2.09 a. u. for S. All electronic structure calculations were performed in the non-magnetic state within DFT, which has been previously shown to appropriately describe the electronic structure of $Co_{1/3}NbS_2$ [9].

$Co_{1/3}NbS_2$ crystal has the hexagonal chiral space group structure $P6_322$. Co atoms are sandwiched between two layers of the parent compound $2H-NbS_2$ [Fig. 1(a)]. A photoemission core level spectrum demonstrates the peaks from p orbitals of Co and Nb atoms, confirming the high quality of the crystals [Fig. 1(b)]. Moreover, the core level spectroscopy was performed on the same surface for ARPES measurement, thus proving the existence of Co atoms after cleaving. We first examine $Co_{1/3}NbS_2$ samples by obtaining the Fermi surface map on the (001) plane with surface sensitive incident photon energy of 81 eV [Fig. 1(c)]. The red hexagon indicates the surface Brillouin zone determined from our experiment. The surface and bulk Brillouin zones of $Co_{1/3}NbS_2$ are also shown in Fig. 1(a) with high symmetry points marked. Notably, the observed hexagonal symmetry from ARPES data doesn't agree with the expected rectangular surface Brillouin zone from a collinear AFM state [24][25]. Therefore, we see no signature of the electronic reconstruction in the AFM state. One possibility is the Co electrons within the first several unit cells on the surface have an itinerant nature instead of being localized. Since ARPES can only probe the first several unit cells, no effect of magnetism can be measured.

To further understand the electronic properties of $Co_{1/3}NbS_2$, we calculated its bulk electronic band structure with and without SOC along high symmetry directions [Figs. 2(a,b)]. Interestingly, there are several Dirac-like crossings near the Fermi level without SOC, as highlighted by the red circles in Fig. 2(a). Moreover, we can also see dispersive bands along Γ-A direction, despite the layered structure of $Co_{1/3}NbS_2$ with long lattice length in $c$. When SOC is included, all the band crossings near the Fermi level are gapped out. However, due to the chiral crystal structure of this material, the screw axis along Γ-A in $Co_{1/3}NbS_2$ protects the doubly degenerate crossing at the high symmetry point Γ [26][27], as indicated by the red circles in Fig. 2(b).

Having established the band structure in $Co_{1/3}NbS_2$, we show the ARPES valence band dispersions along high symmetry directions in Figs. 2(c,d). Calculated band structures with SOC are overlaid on the data. Overall, experimental data show good agreement with calculations. The highly dispersive band [black arrows in Figs. 2 (c,d)] crossing the Fermi level in both directions are well captured by calculations. Specifically, ARPES data show the SOC gapped crossing around 200 meV below the Fermi level [black circle in Fig. 2(b)]. Moreover, the hole pocket (green arrow) at Γ is also observed by ARPES.

Since intuitively $Co_{1/3}NbS_2$ can be treated as the pristine $NbS_2$ doped by the intercalated Co atoms, we compare our experimentally measured $Co_{1/3}NbS_2$ band structure with that of $2H-NbS_2$[28]. Although early studies proposed that intercalation into layers of TMDs would just shift the Fermi level of the bulk band structure (the so called rigid-band-shift picture) [8][20][21], a recent photoemission study on a sister compound suggested that this theory is not valid [29]. Similarly, while both $Co_{1/3}NbS_2$ and $NbS_2$ have pockets around the center and corners of the BZ corresponding to $NbS_2$ [Fig. 1(c) and Figs. SI1(b,d) [30]], a rigid-band-shift picture cannot account for the new pockets at the corners of the red hexagon in Fig. 1(c). Furthermore, below the Fermi level, there is no linear crossing between a hole and an electron band in $NbS_2$ [Fig. SI1(f)], while in the Co intercalated compound, both DFT [black circle in Fig. 2(b)] and ARPES demonstrate the crossings [Figs. 2(c,d)] along high symmetry directions. This crossing can be further confirmed by extracting peaks from the corresponding momentum distribution curves (MDCs) of Figs. 2(c,d) in Fig. SI2. It is apparent that only a rigid band shift cannot produce such crossings in $Co_{1/3}NbS_2$. Therefore, Co intercalation induces a band structure change in a more complex way.

In addition, polarization dependent ARPES data reveals the nature of Co bands near the Fermi level. In our experimental geometry, LH and LV lights mean that the photon polarization is parallel and perpendicular to the scattering plane, respectively, where the scattering plane is normal to the sample surface and lies in the mirror plane of the sample. The analyzer slit direction is also along the mirror plane of the sample. We focus on the d orbitals of Co and Nb, because the polarization dependence reveals the corresponding d orbital nature based on the dipole matrix element effect [31][32]. Specifically, in our experimental geometry, LH (LV) light is sensitive to the initial state with even (odd) parity regarding to the mirror plane. We measured the Fermi surface maps using both LH and LV polarized lights when the analyzer slit is aligned along the Γ-M direction [Figs. 3(a,b)]. Under the LH light, features near the boundary of the Brillouin zone are enhanced compared to the LV light. As a comparison, the bands near the Γ point are of higher intensity with LV light. This can also been seen from the valence band dispersion along Γ-M direction [Figs. 3(c,d)]. Therefore, bands near M (Γ) have an even (odd) orbital nature. Specifically, the bands marked by red and blue arrows in Fig. 3(c) and Fig. SI4(d) are enhanced by LH polarization, so they display an out-of-plane Co $d_{z^2}$ orbital. At the same time, the band indicated by the black arrow in Fig. 3(d) and Fig. SI3(b) is enhanced by LV polarization, thus having an in-plane Co $d_{x^2-y^2}$ orbital. Interestingly, this is opposite to the pristine compound and $Cr_{1/3}NbS_2$ [29], given the experimental geometry is the same for these measurements. Instead of the out-of-plane character near Γ in $NbS_2$ and with Cr intercalation, we find in-plane orbitals at the center in $Co_{1/3}NbS_2$. As a result, it is possible that Co atoms alter the interactions between two $NbS_2$ layers after the intercalation.

It is then important to probe the band dispersion in the c axis direction. Transport study shows that Co atoms serve as the conduction links between $NbS_2$ layers and lead to a much lower resistivity along the c axis compared to the pristine $NbS_2$ compound [33]. Moreover, our DFT calculations reveal dispersive bands along the out-of-plane direction [Fig. 2(a)]. Our data indeed support a stronger dispersion along the c direction, indicating that the quasi-two-dimensional nature of $NbS_2$ band structure is changed. The bulk sensitive ARPES results further exclude the possibility that the new feature (not observed in $NbS_2$) is from the Co atoms on the cleaved surface. We first use soft X-ray incident light to acquire a photon energy dependence of the $\bar{\Gamma}$-$\bar{K}$ valence band cut to study the electronic band structure at various $k_z$ values [Fig. 4(a)]. Clear periodic patterns are observed so that we can determine the positions of high symmetry points (Γ and A), as marked by the red and black lines in Fig. 4(a).

Since we already determine the positions of Γ and A points from photon energy dependence, we can plot the Fermi surface maps corresponding to the two high symmetry points, as shown in Figs. 4(b,c). The pockets at the corners are changed with different $k_z$ positions. While pockets at $\bar{K}$ are connected at the $\bar{M}$ point in Fig. 4(b), they are disconnected in Fig. 4(c), suggesting a strong $k_z$ variation consistent with Fig. 4(a). The pocket around $\bar{\Gamma}$ is also more obvious in Fig. 4(b). Therefore, we observe a clear evolution of band structure in Γ-A direction despite the layered crystal structure of $Co_{1/3}NbS_2$. We also note that, even with surface sensitive VUV ARPES, $Co_{1/3}NbS_2$ already demonstrates a clear three-dimensional dispersion in the out-of-plane direction compared to $NbS_2$ at the Fermi level [Fig. SI5]. Thus, the three-dimensional dispersion of the electronic bands in $NbS_2$ is weak enough that the $k_z$-broadening in VUV completely quenches the ARPES dispersions, while in $Co_{1/3}NbS_2$ the bulk bands are much more pronounced so that the $k_z$-broadening cannot smear them. Therefore, the Co atoms greatly increase the interactions along $k_z$ direction.

As we performed both surface and bulk sensitive ARPES on the Γ-M-K plane [Fig. 1(c) and Fig. 4(b)], it is natural to compare the band structure along the same high symmetry direction. We repeat the same surface sensitive valence band cut from Fig. 2(c) in Fig. 4(d), yet with photon energy in the soft X-ray range. Thanks to the enhanced probe depth and sharp $k_z$ definition [34], the soft X-ray incident light increases the bulk sensitivity of the ARPES experiment and thus should reveal the intrinsic electronic structure of bulk $Co_{1/3}NbS_2$. The corresponding valence band dispersion along Γ-M [Fig. 4(d)] is similar to the surface sensitive result [Fig. 4(e)], with several linear bands crossing the Fermi level indicated by the red and black arrows in Fig. 4(d). However, the linear band marked with the black arrow is split into two lines in Fig. 4(e). In fact, only the inner branch is captured by our bulk band calculations [Fig. 2(c)]. Therefore, DFT results agree with bulk sensitive photoemission results, whereas VUV data demonstrate a band splitting, especially in the second Brillouin zone. This can also be seen from the Fermi surface map [Fig. 1(c)], which shows two circles around $\bar{\Gamma}$ in the second Brillouin zone. On the other hand, only one circle is shown at the same momentum position from soft X-ray data [Figs. 4(b,c)]. Such a difference between VUV and soft X-ray ARPES indicates the band structure on the surface is slightly different from that in the bulk, possibly due to distinct environments at the surface or the relaxation of the cleaved surface. We also confirm that the non-observation of the splitting in soft X-ray ARPES is not due to its lower momentum resolution [Fig. SI6].

In conclusion, using a combination of ARPES and *ab initio* calculations, we map the band structure in $Co_{1/3}NbS_2$. We demonstrate a clear $k_z$ dispersion and show that Co intercalation doesn't simply fit into the rigid-band-shift picture. Our work clearly points out how Co intercalation increases interactions in the out-of-plane direction by altering the quasi-two-dimensional electronic bands in the pristine $NbS_2$ to hosting a three-dimensional nature after intercalation. Finally, a comparison between low and high photon energies suggests the surface environment is different from the bulk, thus inducing a pair of split bands on the surface. Given the layered structure of this compound, and the large AHE value per layer in the exfoliated samples [19], it is necessary to perform nano-ARPES on the exfoliated crystals in the future to understand how bands would change in the two-dimensional limit. Moreover, as suggested by a recent work [35], it is possible to realize a tunable AHE system in $Co_{1/3}NbS_2$ with various doping. Based on our APRES and theoretical calculations, the Fermi surface consists of a hole pocket at Γ and an electron pocket at K in $Co_{1/3}NbS_2$. Furthermore, the size of the hole pocket is slightly larger than that of the pocket at the corner. This resembles the doped Fe-based superconductors, which have a near-perfect nesting between hole and electron pockets at the center and the corner with almost the same size [36]. SDW instability is proposed to arise from this nesting. Here in $Co_{1/3}NbS_2$, if Fermi level is increased by doping, the size of the two pockets will eventually become the same, thus making it possible to realize SDW instability in this system, which would largely enrich the phase diagram of intercalated TMDs. Furthermore, there are several linear crossings [red circles in Fig. 2(a)] near the Fermi level that are gapped by SOC. A large Berry curvature could form due to these SOC induced gaps like some Kagome compounds [37][38]. As a result, they might be the source of the AHE in this compound.

After submission of this work, we found a similar photoemission work [39] that focuses on how Co intercalation doesn't fit into the rigid-band-shift picture from the pristine compound. However, here we extend this point by demonstrating a quasi-two-dimensional to three-dimensional band evolution induced by Co intercalation.

Work at Princeton University and Princeton-led synchrotron based ARPES measurements are supported by the United States Department of Energy (US DOE) under the Basic Energy Sciences program (grant number DOE/BES DE-FG-02-05ER46200). Crystal growth and properties characterization work at George Mason University is supported by the U.S. Department of Energy, Office of Science, Basic Energy Sciences, Materials Science and Engineering Division. Theoretical calculations are supported by NSF Grant No. DMR 1904716. The authors thank the MAX IV Laboratory for access to the Bloch Beamline. We acknowledge Diamond Light Source for time on beamline i05. We acknowledge the Paul Scherrer Institut, Villigen, Switzerland for provision of synchrotron radiation beam time at the ADRESS beam line of the Swiss Light Source [40][41]. We thank C. Cacho and T. Kim for support at beamline i05 of Diamond Light Source. The authors also thank B. Thiagarajan, C. Polley, H. Fedderwitz and J. Adell for beamtime support at Bloch. T.A.C. acknowledges the support of the National Science Foundation Graduate Research Fellowship Program (DGE-1656466). I.B. acknowledges the generous support of the Special Postdoctoral Researchers Program, RIKEN during the late stages of this work. M.Z.H. acknowledges support from Lawrence Berkeley National Laboratory and the Miller Institute of Basic Research in Science at the University of California, Berkeley in the form of a Visiting Miller Professorship.

* Equal contributions
† Corresponding authors: xiany@princeton.edu; mzhasan@princeton.edu

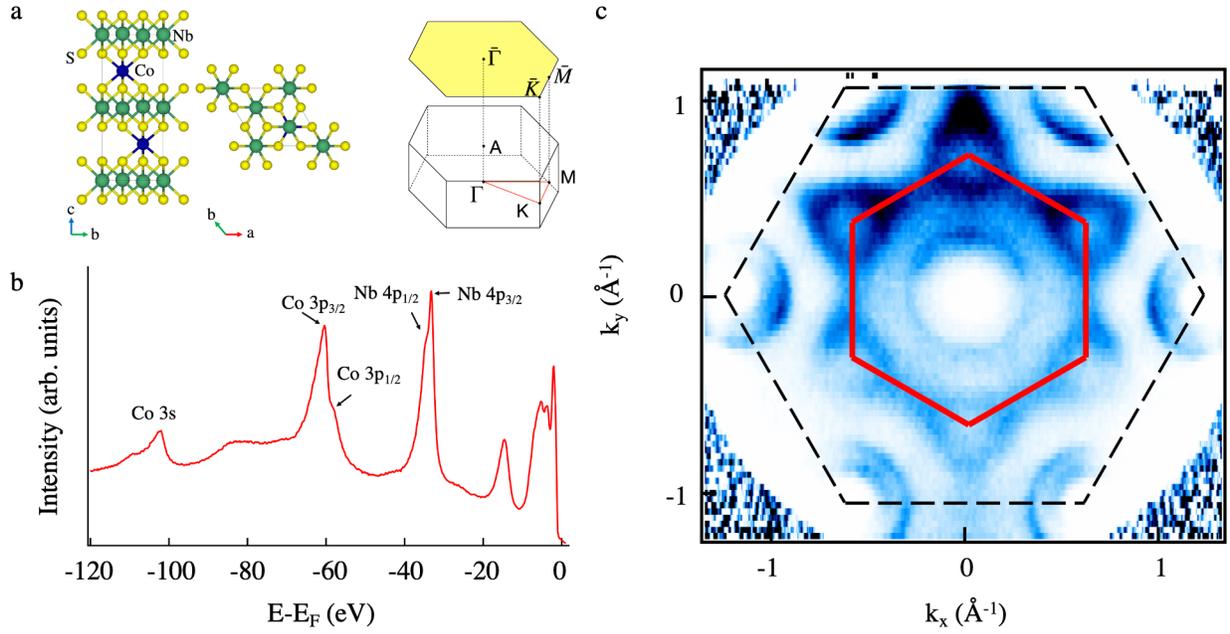

Figure 1. Sample characterization and Fermi surface spectrum of $Co_{1/3}NbS_2$. a. Side view (left) and top view (middle) of the crystal structure of $Co_{1/3}NbS_2$. Right: Bulk and surface Brillouin zones of $Co_{1/3}NbS_2$. High symmetry points are marked. b. Photoemission core level spectrum confirming the high quality of $Co_{1/3}NbS_2$ single crystals. Co and Nb core levels can be identified. c. ARPES Fermi surface map on the (001) plane. The red hexagon represents the first Brillouin zone of $Co_{1/3}NbS_2$ determined from experimental data. Black dotted lines represent the first Brillouin zone of $NbS_2$.

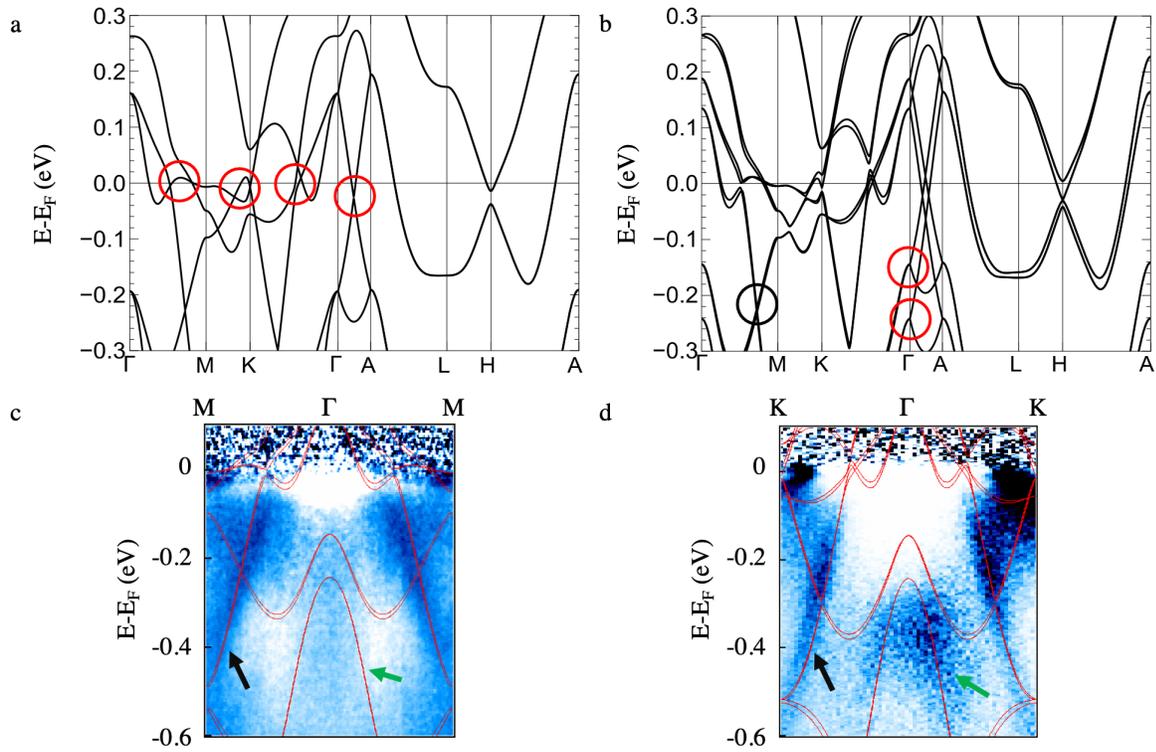

Figure 2. Bulk band structure of $Co_{1/3}NbS_2$. a-b. Theoretical calculations of bulk electronic band structure in $Co_{1/3}NbS_2$ along the high symmetry directions without (a) and with (b) spin-orbit coupling. c-d. ARPES valence band dispersion along Γ-M and Γ-K directions, respectively. DFT calculations are overlaid on top of the data.

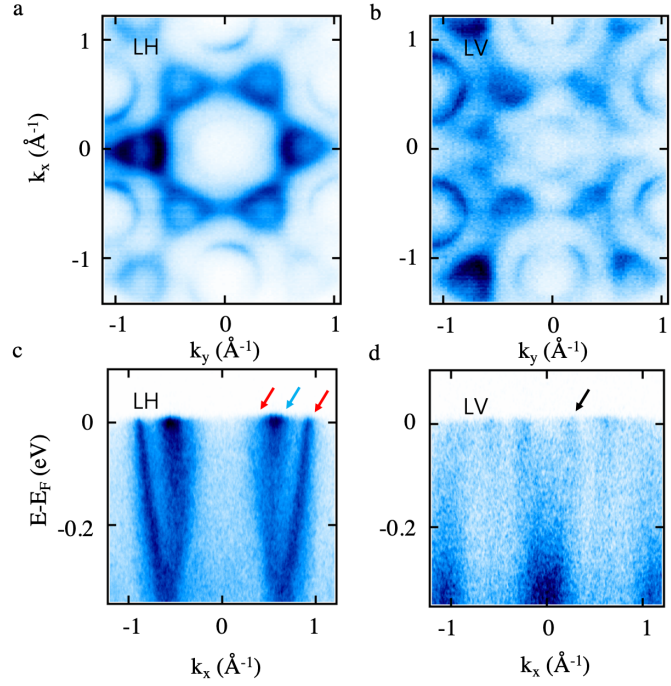

Figure 3. Polarization dependent ARPES data with distinct d orbital characters. a,b. Fermi surface maps in the $k_x - k_y$ plane measured with LH and LV polarized lights. c,d. ARPES valence band dispersion cuts along Γ-M measured with LH and LV polarized lights. Red and blue (black) arrows indicate bands with out-of-plane (in-plane) Co orbitals.

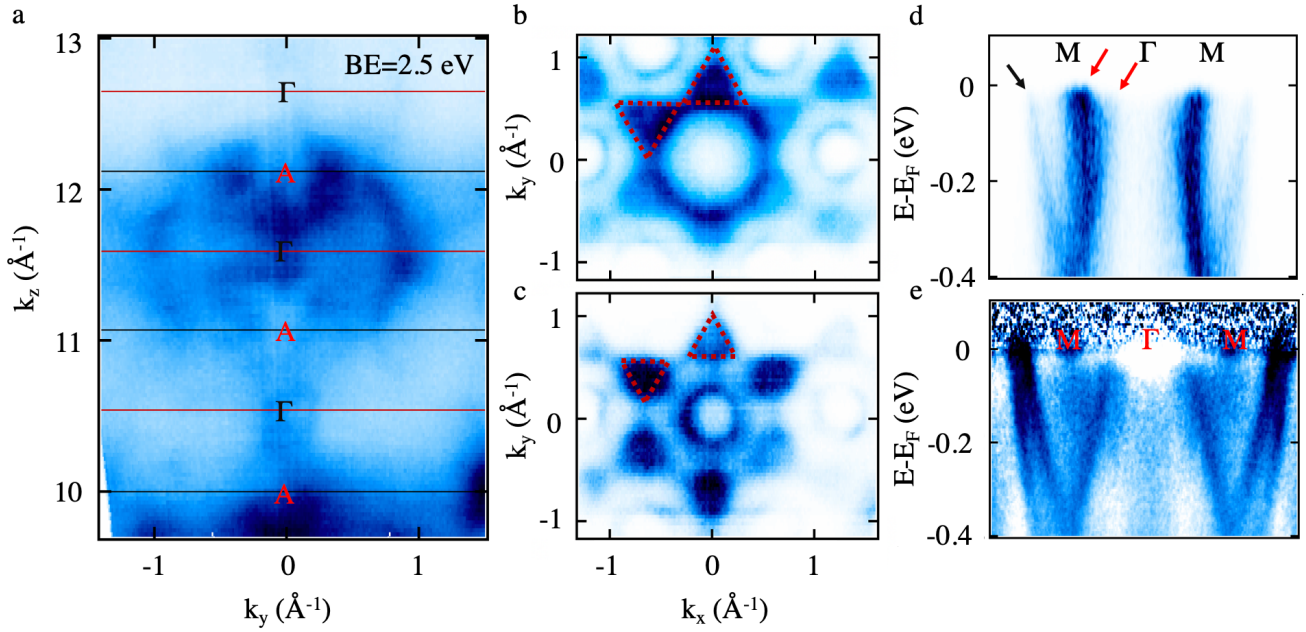

Figure 4. Strong out-of-plane dispersion after Co intercalation. a. Constant energy map in the $k_y - k_z$ plane at the binding energy of 2.5 eV. High symmetry energies are marked by the red and black lines, as periodic patterns along $k_z$ can be resolved. b-c. Fermi surface maps in the $k_x - k_y$ plane with incident photon energy of 413 and 370 eV corresponding to the red and black line in a, respectively. Dotted red triangles are guides to eyes. d-e. ARPES valence band dispersion along Γ-M with bulk and surface sensitivity, respectively.